\documentclass[aps,prd,10pt,preprint,superscriptaddress,onecolumn,nofootinbib]{revtex4-1}
\usepackage{graphicx, epsfig} 
\usepackage{amsmath,amssymb,amsfonts,dsfont,mathrsfs,amsthm,mathtools}
\usepackage{bm} 
\usepackage{color}
\usepackage[usenames]{xcolor}
\usepackage{hyperref}
\usepackage{siunitx}
\hypersetup{linktocpage=blue,colorlinks=true,urlcolor=blue,linkcolor=magenta,citecolor=blue,filecolor=blue}
\usepackage[normalem]{ulem}

\newcommand{\stkout}[1]{\ifmmode\text{\sout{\ensuremath{#1}}}\else\sout{#1}\fi}

\newcommand{\diff}[1]{\text{d}#1}

\newcommand{\Lie}{\mathcal{L}}

\begin{document}



\title{Self-dual Gravitational Instantons in Conformal Gravity: Conserved Charges and Thermodynamics}

\author{Crist\'obal Corral}
\email{crcorral@unap.cl}
\affiliation{Instituto de Ciencias Exactas y Naturales, Facultad de Ciencias, Universidad Arturo Prat, Avenida Arturo Prat Chac\'on 2120, 1110939, Iquique, Chile}

\author{Gast\'on Giribet}
\email{gaston@df.uba.ar}
\affiliation{Physics Department, University of Buenos Aires and IFIBA-CONICET, Ciudad Universitaria, Pabell\'on 1 (1428), Buenos Aires, Argentina}

\author{Rodrigo Olea}
\email{rodrigo.olea@unab.cl}
\affiliation{Universidad Andres Bello, Departamento de Ciencias F\'isicas, \\  Facultad de Ciencias Exactas, Sazi\'e 2212, Piso 7, Santiago, Chile}

\begin{abstract}
We study non-Einstein Bach-flat gravitational instanton solutions that can be regarded as the generalization of the Taub-NUT/Bolt and Eguchi-Hanson solutions of Einstein gravity to conformal gravity. These solutions include non-Einstein spaces which are either asymptotically locally flat spacetimes (ALF) or asymptotically locally Anti-de Sitter (AlAdS). Nevertheless, solutions with different asymptotic conditions exist: we find geometries that present a weakened AlAdS asymptotia, exhibiting the typical low decaying mode of conformal gravity. This permits to identify the simple Neumann boundary condition that, as it happens in the asymptotically AdS sector, selects the Einstein solution out of the solutions of conformal gravity. All the geometries present non-vanishing Hirzebruch signature and Euler characteristic, being single-centered instantons. We compute the topological charges as well as the Noether charges of the Taub-NUT/Bolt and Eguchi-Hanson spacetimes, which happen to be finite. This enables us to study the thermodynamic properties of these geometries. 
\end{abstract}

\maketitle

\section{Introduction}

Conformal gravity is a very interesting theory which has been considered in many different contexts. In four dimensions, the theory is defined by the action whose Lagrangian is the squared Weyl tensor---see Eq.~\eqref{ICG} below---, while in other dimensions polynomial actions with conformal symmetry acquire more involved forms. In eight dimensions, for example, a particular model of conformal gravity can be achieved by simply squaring the four-dimensional Lagrangian, but other 8-dimensional conformal Lagrangians involving specific contractions of four Riemann tensors are possible.\footnote{For a complete classification of conformal invariants in 8 dimensions, see \cite{Boulanger:2004zf}.} In six dimensions, there exist three independent conformal invariant Lagrangians, and in three dimensions, which is another case of interest, a conformal model is possible in terms of the Chern-Simons action for the affine connection.

Among all the contexts in which four-dimensional conformal gravity has been considered, we find that it has been studied as an ultraviolet completion of Einstein gravity \cite{Adler:1982ri}, although the theory is anomalous as its gauge symmetry gets broken by loop corrections. Conformal gravity has also been shown to appear as counterterms in the context of holographic renormalization~\cite{Balasubramanian:2000pq, Anderson2020, Henningson:1998gx, Sen:2012fc}, and applications of AdS/CFT in which conformal gravity represents the bulk theory itself have been considered. Conformal gravity also admits supergravity extensions~\cite{Bergshoeff:1980is, Bergshoeff:1982az, Fradkin:1985am, Liu:1998bu, Ferrara:2018wqd,DAuria:2021dth}, as well as extensions to higher-spin theories~\cite{Fradkin:1985am,Fradkin:1989yd,Fradkin:1989xt,Fradkin:1989md,Nutma:2014pua, Adamo:2016ple, Basile:2018eac}, and topological theories~\cite{Witten:1988xi, Perry:1992ta}. In Ref.~\cite{Berkovits:2004jj}, Berkovits and Witten have shown that conformal gravity appears in the context of string theory on twistor spaces; see also~\cite{Ahn:2004ua, Dolan:2008gc, Adamo:2013tja, Uvarov:2014lfa, Haehnel:2016mlb}. In Ref.~\cite{Maldacena:2011mk}, Maldacena has observed a remarkable connection between classical four-dimensional conformal gravity and Einstein gravity and used it to show that the former theory, when supplemented with an appropriate boundary condition, led to obtain the semiclassical wavefunction of the universe of asymptotically de Sitter spacetime. This is achieved by removing the ghost mode that conformal gravity exhibits by imposing the mentioned boundary condition. This observation gave rise to an unfathomable series of research articles revisiting conformal gravity and its extensions. An incomplete list of recent developments is \cite{Edery:1998zi,Lu:2011zk, Deser:2011xc, Lu:2011ks, Anninos:2011jp, Lu:2011mw, Mannheim:2011is, Bergshoeff:2012sc, Deser:2012qg, Lu:2013hx, Dunajski:2013zta, Wheeler:2013ora, Grumiller:2013mxa, Alishahiha:2013dca, Aros:2013yaa, Miskovic:2014zja, Beccaria:2015uta, Anastasiou:2016jix, Chernicoff:2018apt, Anastasiou:2021tlv, Irakleidou:2016xot, Liu:2012xn,Momennia:2019edt,Momennia:2019cfd,Momennia:2021aca}; see also references therein.

Not many \emph{bona fide} interesting solutions to conformal gravity field equations are known. Among the most {remarkable} ones, there are non-Einstein spaces that include static and stationary black holes \cite{Riegert:1984zz, Liu:2012xn, Lu:2012xu} with different asymptotics, and gravitational waves \cite{Gullu:2011sj, AyonBeato:2012da}. The theory, however, contains a richer variety of inequivalent sectors. Here, we will study gravitational instanton solutions to four-dimensional conformal gravity. More precisely, we will discuss a generalization of the Euclidean Taub-NUT~\cite{Taub:1950ez,Newman:1963yy} and Taub-Bolt~\cite{Page:1979aj} solutions to the case of conformal gravity.\footnote{In presence of nonminimally coupled scalar fields respecting conformal invariance, some Taub-NUT solutions have been found in Refs.~\cite{Bardoux:2013swa,Cisterna:2021xxq}.} This family of solutions presents self-dual geometries and includes the most general static, spherically symmetric black hole solutions as particular cases. This {comprehends} both asymptotically locally flat and asymptotically locally AdS geometries, and so it extends previous studies of Bach-flat {spaces} with NUT charge. We compute the mass and entropy of the solutions using the Noether-Wald formalism~\cite{Wald:1993nt,Iyer:1994ys} showing that they are finite without any reference to counterterms. The Euclidean on-shell action is also obtained and we show that the thermodynamic variables satisfy the Gibbs-Duhem relation and the first law. The solutions also have non-vanishing topological charges. 

The paper is organized as follows: In Sec.~\ref{sec:CG}, we review conformal gravity in four dimensions. The generalization of Taub-NUT and Taub-Bolt solutions is discussed in Sec.~\ref{sec:BFTN}. We compute the conserved charges of the solutions in Sec.~\ref{sec:charges}. In Sec.~\ref{sec:comparison}, we compare the solutions with other instantons found in the literature~\cite{Strominger:1984zy, IWAI199355, Miyake1995}. In Sec.~\ref{sec:eguchi-hanson}, we make some comments about other self-dual Bach-flat solutions that can be constructed, presenting Bach-flat generalizations of the Eguchi-Hanson gravitational instanton. Finally, in Sec.~\ref{sec:conclusion} we present our conclusions.

\section{Conformal gravity in four dimensions\label{sec:CG}}

The action principle of conformal gravity in four spacetime dimensions is constructed out of the simplest non-topological conformal invariant in such dimensionality: the Weyl squared term. Thus, it can be expressed as
\begin{align}\label{ICG}
    I_{\rm CG}[g_{\mu\nu}] &= \alpha_{\rm CG}\int_{\mathcal{M}}\diff{^4x}\sqrt{-g}\,W_{\mu\nu}^{\lambda\rho}W^{\mu\nu}_{\lambda\rho} \,,
\end{align}
where $\alpha_{\rm CG}$ is the coupling constant of the theory and the Weyl tensor is defined as 
\begin{align}\label{Weyl}
W^{\mu\nu}_{\lambda\rho} &=\frac{1}{4} \left(\delta_{\sigma\tau}^{\mu\nu}\delta_{\lambda\rho}^{\gamma\delta} - 8 \delta_{[\lambda}^{[\delta}\delta_{\rho]}^{[\mu}\delta_{[\sigma}^{\nu]}\delta_{\tau]}^{\gamma]} + \frac{1}{3}\delta_{\lambda\rho}^{\mu\nu}\delta_{\sigma\tau}^{\gamma\delta} \right)  R^{\sigma\tau}_{\gamma\delta} \equiv \left(\Xi^{\mu\nu}_{\lambda\rho}\right)^{\gamma\delta}_{\sigma\tau} R^{\sigma\tau}_{\gamma\delta} \,.
\end{align}
Written in this way, it becomes clear that $\Xi$ projects the Riemann tensor onto their completely traceless irreducible component. Indeed, since the Weyl tensor is already traceless in any of their indices, their contraction with $\Xi$ projects the latter onto itself, and it is therefore straightforward to show the idempotency of $\Xi$.

The field equations are obtained by performing stationary variations of the action~\eqref{ICG} with respect to the metric, giving (see for example~\cite{Anastasiou:2016jix})
\begin{align}
    \delta I_{\rm CG} = \alpha_{\rm CG}\int_{\mathcal{M}}\diff{^4 x} \sqrt{-g}\, \delta g^{\mu\nu}B_{\mu\nu} + \alpha_{\rm CG}\int_{\mathcal{M}}\diff{^4x}\sqrt{-g}\,\nabla_\mu\Theta^\mu\,,
\end{align}
where $B_{\mu\nu}$ and $\Theta^\mu$ are the Bach tensor and the boundary term arising from the variation, respectively; explicitly, 
\begin{align}\label{eom}
 B_{\mu\nu} &= \nabla^\lambda C_{\mu\nu\lambda} - S^{\lambda\rho} W_{\lambda\mu\nu\rho}\,, \\
\label{bt}
 \Theta^\mu &= 4\delta\Gamma^{\lambda}{}_{\nu\rho} W_{\lambda}{}^{\rho\mu\nu} - 4\delta g_{\nu\sigma} \nabla_\rho W^{\nu\mu\rho\sigma} = -4\nabla_\rho\delta g_{\nu\sigma} W^{\rho\sigma\mu\nu} + 4\delta g_{\nu\sigma}\nabla_\rho W^{\rho\sigma\mu\nu}.  
 \end{align}
Here, $B_{\mu\nu} = 0$ represents the field equations with $C_{\mu\nu\lambda}$ and $S_{\mu\nu}$ being the Cotton and Schouten tensors, defined through
\begin{align}\label{Cotton&Schouten}
    C_{\mu\nu\lambda} =  2\nabla_{[\lambda}S_{\nu]\mu} \;\;\;\;\; \mbox{and} \;\;\;\;\; S_{\mu\nu} = \frac{1}{2}\left(R_{\mu\nu}
    - \frac{1}{6}g_{\mu\nu} R \right),
\end{align}
respectively, where antisymmetrization is normalized as $A_{[\mu\nu]} = \tfrac{1}{2}\left(A_{\mu\nu} - A_{\nu\mu} \right)$. In fact, notice that any Einstein spacetime with $R^\mu_\nu = - \tfrac{3}{\ell^2}\delta^\mu_\nu$ is solution of conformal gravity, since the Schouten tensor becomes $S^\mu_\nu = - \tfrac{1}{2\ell^2}\delta^\mu_\nu$ while the Cotton tensor vanishes identically. These two identities implies that the Bach tensor is identically zero for Einstein spaces. 

There are, however, Bach-flat solutions which are not Einstein spacetimes. In the next section, we find an explicit Taub-NUT/Bolt solution with these properties. 

\section{Bach-flat Taub-NUT/Bolt solution\label{sec:BFTN}}

In order to solve the field equations~\eqref{eom} analytically, we assume the inhomogeneous Euclidean family of metrics proposed in Ref.~\cite{Page:1985bq}. In particular, we focus on the $U(1)$ fibration of $S^2$ described by the line element
\begin{align}\label{ansatz}
    \diff{s^2} = f(r)\left(\diff{\tau} + 2n\cos\theta\, \diff{\phi} \right)^2 + \frac{\diff{r^2}}{f(r)} + \left(r^2-n^2\right)\left(\diff{\theta^2} + \sin^2\theta \,  \diff{\phi^2} \right),
\end{align}
where $\tau$ is the Euclidean time and $n$ is known as the NUT charge, which is related to the first Chern number at infinity~\cite{Hawking:1998ct}. Indeed, the magnetic part of the Weyl tensor is sensitive to the presence of $n$; the latter is usually interpreted as the magnetic mass of the spacetime (see for instance~\cite{RevModPhys.70.427,PhysRevD.93.084022,Flores-Alfonso:2018jra} and references therein). 

We find that the metric function that solves the field equations of conformal gravity is
\begin{align}\label{fsol}
    f(r) &= \frac{2n^2}{3\left(r^2-n^2 \right)} - \frac{m\left(3r^2 + n^2 \right)}{r^2-n^2} + \frac{p r\left(r^2+3n^2 \right)}{r^2-n^2} + \frac{q r\left(3r^2+n^2 \right)}{r^2-n^2} + \frac{\lambda r^2\left(r^2+3n^2 \right)}{r^2-n^2},
\end{align}
where $m,p,q,$ and $\lambda$ are integration constants subject to the condition 
\begin{align}\label{Relation}
    9\lambda^2 n^4-30\lambda m n^2- 9n^2p^2 - 30n^2pq-9n^2q^2-8\lambda n^2 + 9m^2-1 = 0,
\end{align}
which generalizes the Taub-NUT solution of general relativity \cite{Taub:1950ez, Newman:1963yy, Misner:1963fr}. These solutions can likely be included in a Plebanski-Demianski type ansatz \cite{Liu:2012xn}. It might be convenient to rewrite the expression~\eqref{fsol} above as follows
\begin{align}\label{fsol2}
    f(r) &= \frac{r^2+n^2-2\hat{m}r+{\ell}^{-2}(r^4-6n^2r^2-3n^4) +\, a\, (r^2+n^2/3)+\, b\, r^3 }{r^2-n^2 },
\end{align}
where $\hat{m}=-(n^2/2)(3p+q)$, ${\ell}^2=1/\lambda $, $a=9\lambda n^2-3m-1$, and $b=p+3q$. Written in this way, the condition on the parameters in Eq.~\eqref{Relation} becomes
\begin{align}\label{Relation2}
  a^2\ell^2 + 6b\ell^2\hat{m} + 2a\ell^2 - 8an^2 =0.
\end{align}
The asymptotic behavior of the metric function as $r\to\infty$ is
\begin{align}\label{linear}
    f(r) = \frac{r^2}{\ell^2} + br + 1 - \frac{5n^2}{\ell^2} + a + \frac{bn^2-2\hat{m}}{r} + \mathcal{O}(r^{-2}).
\end{align}
As it happens with the spherically symmetric solution in conformal gravity, the constraint $b=0$ has to be imposed for the asymptotic behavior of Einstein gravity to be recovered. This suffices to eliminate the $\mathcal{O}(r)$ term in (\ref{linear}).

Although all Einstein spaces are Bach flat, the reciprocal assertion is not true: there are Bach flat metrics which are non-Einstein. Indeed, the solution of Eq.~\eqref{fsol2} has the latter property as it can be seen from their non-constant Ricci scalar, that is
\begin{align}\label{RicciscalBF}
     R = -\frac{12}{\ell^2} - \frac{2\left(a+3br \right)}{r^2-n^2}.
\end{align}
This shows that this is a 1-parameter deformation of the Einstein space since, recall, $a$ and $b$ are related through Eq.~\eqref{Relation2}. Additionally, since the traceless Ricci tensor $H_{\mu\nu} = R_{\mu\nu} - \tfrac{1}{4}g_{\mu\nu}R$ is identically zero for Einstein spaces, it can be used as an additional test of the non-Einstein nature the solution presented here. Their traceless Ricci squared is
\begin{align}\label{TLR2BF}
    H_{\mu\nu}H^{\mu\nu} 
    = \left[\frac{br\left(r^2+3n^2 \right) + a\left(r^2+n^2/3 \right) }{\left(r^2-n^2 \right)^2}\right]^2\,.
\end{align}

The form for the metric function in Eq.~\eqref{fsol2} permits to analyze the properties of the solution in a simple way. For example, one observes from there that the case $a=b=0$ corresponds to an Einstein manifold; more precisely, it corresponds to the (A)dS-Schwarzschild-Taub-NUT solution to cosmological Einstein equations with mass $\hat{m}$, NUT charge $n$, and (A)dS radius ${\ell}$. The non-Einstein parameter $b$ is associated to a new mode of conformal gravity, to which we will refer to as the $b$-mode. Indeed, the parameter $b=p+3q$ is the one that controls in~\eqref{linear} the linear mode $\mathcal{O}(r)$, which is a characteristic feature of non-Einstein solutions of conformal gravity. This is the mode studied in Ref.~\cite{Maldacena:2011mk} to derive Einstein theory from the conformal gravity by imposing a Neumann boundary condition. The latter precisely realizes the freezing of the linearly growing mode by setting $b=0$. Such linear mode can also be seen to appear in the general static, asymptotically flat black hole solutions of conformal gravity \cite{Riegert:1984zz}; see~\eqref{Riegert3} below. By taking $n=0$ and keeping all other parameters fixed, the solution above reduces to
\begin{equation}\label{Riegert1}
    f(r)=  1 +br+\lambda {r^2}\, ,
\end{equation}
where we used Eq.~\eqref{Relation}. This is one of the static solutions studied by Riegert in \cite{Riegert:1984zz}. We see from Eq.~\eqref{Riegert1} that the solutions with $b\neq 0 $ and $\lambda <0$ exhibit a weakened versions of AdS asymptotic conditions. As said, the Neumann boundary conditions considered in Ref.~\cite{Maldacena:2011mk} corresponds to eliminate the $b$-mode, which is a feature of conformal gravity. The same feature can be found, for example, in 3-dimensional massive gravity, 3-dimensional conformal gravity, and in models of conformal gravity in higher dimensions.

Another special case of the Riegert solution is found when performing a different limit: Consider $p=-3k/(2n^2)$ and $q=k/(2n^2)$, and then take the limit $n\to 0$. This yields
\begin{equation}\label{Riegert2}
    f(r)=  1 -\frac{4k}{r}+\lambda {r^2}\, ,
\end{equation}
where, again, we used Eq.~\eqref{Relation}. This is nothing but the (A)dS-Schwarzschild black hole solution, for which $a=b=0$. This is, of course, one of the branches of the spherically symmetric Bach-flat solution. The most general spherically symmetric case is obtained by considering $p=-b/2-3k/(2n^2) $, $q=b/2+k/(2n^2) $, which in the limit $n\to 0$ yields
\begin{equation}\label{Riegert3}
    f(r)= -3m + b r -\frac{4k}{r}+\lambda {r^2}\, ,
\end{equation}
where the constants are related by
\begin{equation}\label{Riegert4}
    9m^2-1+12bk=0.
\end{equation}
Indeed, Eqs.~\eqref{Riegert3}-\eqref{Riegert4} turns out to be the most general static, spherically symmetric solution to conformal gravity field equations \cite{Riegert:1984zz, Mannheim:1988dj}. In other words, the gravitational instanton with symmetry $SO(3)\times \mathbb{R}$ that is obtained by considering (\ref{Riegert3}) with $n=0$ in (\ref{ansatz}) is nothing but the Euclidean version of the Riegert black hole of conformal gravity.



\subsection{Nuts and bolts}

Regularity of Euclidean hypersurfaces with either zero or two-dimensional fixed points defines the conditions for nuts and bolts, respectively; they are
\begin{align}\label{nuts}
    &\mbox{NUT:}& f(n) &=0 & &\mbox{and}& f'(r)\Big|_{r=n} &= \frac{4\pi}{\beta_{\rm nut}},\\
    \label{bolts}
    &\mbox{Bolt:}& f(r_b) &=0 & &\mbox{and}& f'(r)\Big|_{r=r_b} &= \frac{4\pi}{\beta_{\rm bolt}}.
\end{align}
Here, $\beta_{\rm nut}$ and $\beta_{\rm bolt}$ is the period of the Euclidean time to avoid the presence of conical singularities in each case, and $r_b>n$ is the bolt radius as defined below [see Eq.~\eqref{mbolt}]. For Taub-NUT, Eq.~\eqref{nuts} implies 
 \begin{align}\label{mqnut}
\hat{m} &= n + \frac{2an}{3} + \frac{bn^2}{2} - \frac{4n^3}{\ell^2} \equiv \hat{m}_{\rm nut} \;\;\;\;\; \mbox{and} \;\;\;\;\;   a = -3bn \equiv a_{\rm nut},
 \end{align}
 which solves automatically the relation~\eqref{Relation2}. For Taub-Bolt, on the other hand, condition~\eqref{bolts} fixes
 \begin{align}\label{mbolt}
      \hat{m} &= \frac{r_b^2+n^2}{2r_b} + \frac{a\left(3r_b^2+n^2\right)}{6r_b} + \frac{br_b^2}{2} + \frac{r_b^4-6n^2r_b^2-3n^4}{2r_b\ell^2}  \equiv \hat{m}_{\rm bolt}\,.
 \end{align}
This polynomial equation of degree four defines implicitly the bolt radius $r_b$. Using these relations, the explicit form of the metric functions are
\begin{align}\label{fnut}
   f_{\rm nut}(r) &=\frac{r-n}{r+n} + \frac{b\left(r-n\right)^2}{r+n} + \frac{(r-n)^2(r+3n)}{\ell^2(r+n)}\,, \\
    \notag
   f_{\rm bolt}(r) &= \frac{(r-r_b)(rr_b-n^2)}{r_b\left(r^2-n^2\right)} + \frac{a(r-r_b)(3rr_b-n^2)}{3r_b\left(r^2-n^2 \right)} + \frac{br\left(r^2-r_b^2 \right)}{r^2-n^2}  \\
   \label{fbolt}
    &\quad + \frac{\left(r-r_b\right)\left(3n^4 - 6 n^2 r r_b + r^3 r_b + r^2r_b^2 + r r_b^3\right)}{r_b\ell^2\left(r^2-n^2 \right)}\,.
\end{align}
Thus, for the Bach-flat Taub-NUT and Taub-Bolt metrics presented above, the curvature invariants become finite and the manifold is completely regular, provided that the Euclidean time is periodically identified. In particular, the Ricci scalar for Taub-NUT [see Eq.~\eqref{fnut}] becomes
\begin{align}
    R &=  - \frac{12}{\ell^2}  - \frac{6b}{r+n}\,,
\end{align}
while, for Taub-Bolt, the Ricci scalar remains as in Eq.~\eqref{RicciscalBF} with the condition $r\geq r_{b}>n$ guaranteeing regularity. 

The period of the Euclidean time for Taub-NUT can be obtained from the second condition in Eq.~\eqref{nuts} giving $\beta_{\rm nut}=8\pi n$. For Taub-Bolt, on the other hand, the period of the Euclidean time is found to be
\begin{align}
    \beta_{\rm bolt} &= \frac{12\pi r_b \left(r_b^2-n^2\right)}{6br_b^3 + a\left(3r_b^2-n^2 \right) + 3\left(r_b^2-n^2\right)  +\frac{9}{\ell^2}\left(r_b^2-n^2 \right)^2}.
\end{align}
Additionally, unobservability of the position of the Misner string imposes that the period of the Euclidean time for Taub-Bolt must be equal to the one of Taub-NUT, i.e. $\beta_{\rm bolt}=8\pi n$. This condition, in turn, implies that both solutions have the same temperature. Indeed, for Einstein spaces, the unobservability of the Misner string leaves only one free parameter: the NUT charge. For the Bach-flat solutions studied here, however, the NUT charge and the $b$-mode remain free; up to Neumann boundary conditions that picks up the former as the only independent parameter.

\subsection{Euclidean on-shell action and topological invariants}

It is worth noticing that, for NUT, the Weyl tensor becomes globally (anti-)self dual, i.e.,
\begin{align}\label{Wdual}
 W^{\mu\nu}_{\lambda\rho} = \pm\frac{1}{2}\varepsilon_{\lambda\rho\sigma\tau}W^{\mu\nu\sigma\tau} \equiv \pm \tilde{W}^{\mu\nu}_{\lambda\rho} ,
\end{align}
where $\varepsilon_{\mu\nu\lambda\rho}$ is the Levi-Civita tensor. Indeed, using the off-shell identity
\begin{align}\label{WWdual=RRdual}
\varepsilon^{\mu\nu \lambda\rho }W_{\sigma\tau\lambda\rho}W^{\sigma\tau}_{\mu\nu }=\varepsilon^{\mu\nu \lambda\rho }R_{\sigma\tau\lambda\rho}R^{\sigma\tau}_{\mu\nu }\,,    
\end{align}
the Euclidean on-shell action for Taub-NUT can be computed directly in terms of a topological invariant, namely,
\begin{align}\label{Inut}
    I_{\rm nut} &= \pm \frac{\alpha_{\rm CG}}{2}\int_{\mathcal{M}}\diff{^4x}\sqrt{g}\,\varepsilon^{\mu\nu \lambda\rho }W_{\sigma\tau\lambda\rho}W^{\sigma\tau}_{\mu\nu } = \pm 16\pi^2\alpha_{\rm CG}\,c\,,
\end{align}
where the Chern-Pontryagin index is defined as
\begin{align}
c &= \frac{1}{32\pi^2} \int_{\mathcal{M}} d^4x\,\sqrt{g}\, \varepsilon^{\mu\nu \lambda\rho }R_{\sigma\tau\lambda\rho}R^{\sigma\tau}_{\mu\nu }= 2 - \frac{16n^2}{\ell^2}\left(1-\frac{2n^2}{\ell^2}\right) + 4b^2n^2 \, ,\label{signatura}
\end{align}
and it enters in the topological invariant that measures the difference between harmonic self-dual and anti self-dual forms on the manifold, i.e., the Hirzebruch signature (see, for instance, Ref.~\cite{Eguchi:1980jx}). The solution becomes anti-self dual for $n<0$ and the right-hand side of Eq.~\eqref{signatura} picks a global minus sign. Thus, the Euclidean on-shell action of conformal gravity evaluated on the Taub-NUT solution~\eqref{fnut} saturates the BPS bound and it is proportional to the Chern-Pontryagin index, similar to instantons in Yang-Mills theory. The latter is related to the index of the Dirac operator and it  is related to the axial anomaly in the quantum theory~\cite{AlvarezGaume:1983ig,AlvarezGaume:1984dr}. Additionally, notice that the massive mode contributes to the Chern-Pontryagin index, as it can be checked from the last term in Eq.~\eqref{signatura}.

One could add topological invariants to the conformal gravity action~\eqref{ICG} without affecting the bulk dynamics. Introducing the Chern-Pontryagin invariant, for instance, yields 
\begin{align}\label{ICGPont}
    I_{\rm CGP} = \alpha_{\rm CG}\int_{\mathcal{M}}\diff{^4x}\sqrt{g}\;W^{\mu\nu}_{\lambda\rho}W_{\mu\nu}^{\lambda\rho} +\frac{\vartheta}{32\pi^2} \int_{\mathcal{M}} d^4x\,\sqrt{g}\, \varepsilon^{\mu\nu \lambda\rho }R_{\sigma\tau\lambda\rho}R^{\sigma\tau}_{\mu\nu }. 
\end{align}
The tensor $\tilde{W}^{\, \nu }_{\mu \, \lambda\rho}$ is invariant under Weyl transformations and so the Chern-Pontryagin index. Indeed, by setting
\begin{align}\label{Belucci}
    \vartheta = \pm 16\pi^2\alpha_{\rm CG}\,,
\end{align}
and using the off-shell identity~\eqref{WWdual=RRdual},
the action principle~\eqref{ICGPont} can be written as
\begin{align}
  I_{\rm CGP} =  \frac{\alpha_{\rm CG}}{2}\int_{\mathcal{M}} \diff{^4x}\sqrt{g}\,\Big(W^{\mu \nu}_{\lambda \rho }\pm \tilde{W}^{\mu \nu}_{\lambda \rho }\Big) \Big(W_{\mu \nu}^{\lambda \rho }\pm \tilde{W}_{\mu \nu}^{ \lambda \rho }\Big). 
\end{align}
The Euclidean action of conformal gravity, augmented by the topological invariant of Eq.~\eqref{signatura}, becomes identically zero when evaluated at the (anti-)self dual Taub-NUT. Thus, the role of the Chern-Pontryagin invariant in Eq.~\eqref{ICGPont} is to set the whole family of (anti-)self dual Taub-NUT solutions as the ground state of conformal gravity, similar to what happens in four-dimensional general relativity enhanced by the presence of topological terms (see Refs.~\cite{Miskovic:2009bm,PhysRevD.93.084022,Ciambelli:2020qny}).

The other topological invariant is given by the Euler characteristic, $\chi$, namely
\begin{align}
 \chi &=\frac{1}{32\pi^2}\left[\int_{\mathcal{M}} d^4x\,\mathcal{G} + \int_{\partial\mathcal{M}}\diff{^3x}\; \mathcal{B}\right] = 1\,, \label{eulerch}
\end{align}
where 
\begin{align}
    \mathcal{G} &= \sqrt{g} \left( R_{\lambda\rho \sigma\tau}R^{\lambda\rho \sigma\tau}-4R_{\lambda\rho }R^{\lambda\rho }+R^2\right)\,,\\
    \mathcal{B} &= 4\sqrt{h}\;\delta^{\alpha\beta\gamma}_{\mu\nu\lambda}K^\mu_\alpha\left(\frac{1}{2}\mathcal{R}^{\nu\lambda}_{\beta\gamma}(h) - \frac{1}{3}K^\nu_\beta K^\lambda_\gamma \right),
\end{align}
are the Gauss--Bonnet and 2nd Chern form, respectively. Here, $h_{\mu\nu} = g_{\mu\nu} - n_\mu n_\nu$ is the induced metric on the asymptotic boundary with $h$ being its determinant, $n^\mu$ denotes its normal unit vector, $\mathcal{R}^{\mu\nu}_{\lambda\rho}(h)$ is the intrinsic Riemann curvature associated to $h_{\mu\nu}$ and $K_{\mu\nu}= h^{\lambda}{}_\mu\nabla_\lambda n_\nu $ is the extrinsic curvature. This is consistent with the indices of the gravitational instantons of type $A_k$ \cite{Hawking:1976jb, Gibbons:1979zt}, which describe multi-Taub-NUT solutions with $k+1$ centers. The Euler characteristic for these solutions is $k+1$, yielding $\chi=1$ for the Taub-NUT metric, which corresponds to $k=0$. Notice that the Euler characteristic (\ref{eulerch}) does not depend on the value of the additional parameters such as $b$.

The Taub-Bolt solution is no longer globally self dual as Taub-NUT was. However, the Euclidean on-shell action is finite without need of adding counterterms. Explicitly, the Bach-flat Taub-Bolt solution in Eq.~\eqref{fbolt} yields
\begin{align}\notag
    I_{\rm Bolt} &= 128\pi^2\alpha_{\rm CG}n\Bigg\{\frac{ a \left(n^2-3r_b^2\right)}{9\left(r_b^2-n^2\right)}\left[\frac{ar_b + 3b\left(r_b^2+n^2 \right)}{r_b^2-n^2} + \frac{3\left(r_b^2-n^2+\ell^2 \right)}{\ell^2r_b} \right] \\
    & -  \frac{b^2r_b^3\left(r_b^2+n^2 \right)}{\left(r_b^2 -n^2\right)^2} - \frac{2br_b^2\left(r_b^2-n^2+\ell^2 \right)}{\ell^2\left(r_b^2-n^2 \right)} - \frac{r_b^4+3n^4+\ell^4-4\ell^2n^2+2\ell^2 r_b^2}{\ell^4 r_b} \Bigg\}\,.
\end{align}

From a holographic viewpoint, conformal gravity, as a higher-derivative theory it is, brings in new boundary sources along with correlators associated to them \cite{Grumiller:2013mxa}. It would be important to provide a more formal proof on the finiteness of the action for non-Einstein solutions, at least, in the weakly decaying, asymptotically AdS sector. As a matter of fact, taking the standard asymptotic falloff of the curvature substantially restricts the type of sub-leading deformations in the metric, respect to the Einstein branch of the theory \cite{Ghodsi:2014hua,Anastasiou:2017mag}.

\section{Conserved Noether Charges and Wald's formula\label{sec:charges}}

In this section, we use the Noether-Wald formalism~\cite{Wald:1993nt,Iyer:1994ys} to compute the conserved charges associated to the Bach-flat Taub-NUT/Bolt solutions. This can be used to obtain the mass and entropy of the solution as shown next.

Diffeomorphism invariance of the action~\eqref{ICG} leads to the identity
\begin{align}\label{conservationlaw}
 \nabla_\mu J^\mu = -\Lie_\xi g^{\mu\nu}B_{\mu\nu},
\end{align}
where $\Lie_\xi$ is the Lie derivative along the vector field $\xi=\xi^\mu\partial_\mu$ that generates the diffeomorphism invariance, and the Noether current is defined through
\begin{align}
 J^\mu = -4\alpha_{\rm CG}\nabla_\nu\left(W^{\mu\nu\rho\sigma}\nabla_\rho\xi_\sigma + 2\xi_\rho C^{\rho\mu\nu} \right).
\end{align}
We expressed the last term in terms of the Cotton tensor through $C_{\mu\nu\lambda}=-\nabla_\sigma W^{\sigma}{}_{\mu\nu\lambda}$. Clearly, Eq.~\eqref{conservationlaw} becomes a conservation law for the Noether current when the field equations hold, namely, when $B_{\mu\nu}=0$. The Poincaré lemma, in turn, allows us to express the Noether current as $J^\mu = \nabla_\nu q^{\mu\nu}$, where $q^{\mu\nu} = -q^{\nu\mu}$ is the 2-form Noether prepotential given by
\begin{align}\label{NWprepotential}
    q^{\mu\nu} &= -4\alpha_{\rm CG} \left(W^{\mu\nu\rho\sigma}\nabla_\rho\xi_\sigma + 2\xi_\rho C^{\rho\mu\nu} \right).
\end{align}
If $\xi$ is a Killing vector, the conserved Noether charge is obtained by integrating the Noether prepotential~\eqref{NWprepotential} over a codimension-2 hypersurface $\Sigma$, that is
\begin{align}\label{noethercharge}
 Q[\xi] = \frac{1}{2}\int_\Sigma\epsilon_{\mu\nu\lambda\rho}q^{\mu\nu}\diff{x^{\lambda}}\wedge\diff{x^{\rho}} \equiv \int_\Sigma Q_{\mu\nu}\diff{x^{\mu}}\wedge\diff{x^{\nu}} = \int_\Sigma\mathbf{Q}\,.
\end{align}

Let us focus first on the the Killing vector $\xi=\partial_\tau$ that generates the Euclidean time isometry. In this case, the relevant components of $Q_{\mu\nu}$ to compute the conserved charges are
\begin{align}\notag
 Q_{\theta\phi} &=  \frac{2\alpha_{\rm CG}}{3}\Bigg[ 2\left(r^2-n^2\right)f'''f - \left[\left(r^2-n^2 \right)f'-2fr \right]f'' + 2rf'^2   \\
 &\quad - \frac{2\left[17fn^2 + 3fr^2 - \left(r^2-n^2\right) \right]f'}{r^2-n^2} + \frac{4fr\left[17fn^2 + fr^2 - \left(r^2-n^2\right) \right]}{\left(r^2-n^2 \right)^2}\Bigg]\sin\theta\,, \\
 Q_{r\phi} &= -\frac{8\alpha_{\rm CG}n^2}{3\left(r^2-n^2 \right)}\Bigg[ 2f''f-3f'^2 + \frac{8f'fr}{r^2-n^2} - \frac{4f\left[7fn^2+2fr^2-2\left(r^2-n^2 \right) \right]}{\left(r^2-n^2 \right)^2}\Bigg]\cos\theta\,,
\end{align}
where prime denotes derivative with respect to the radial coordinate. 

The mass of the solution~\eqref{fsol} is obtained by evaluating the Noether-Wald charge generated by $\xi=\partial_\tau$ at the boundary {located at radial infinity}; that is
\begin{align}
   M =  \lim_{r\to\infty}\int_0^{2\pi}\int_0^\pi\diff{\theta}\diff{\phi}\, Q_{\theta\phi} = -16\pi\alpha_{\rm CG}\left[ \lambda n^2\left(p-13q \right) + \left(m+\tfrac{1}{3}\right)\left(p + 3q\right)\right].
\end{align}
This yields
\begin{align}
   M =\frac{16\alpha_{\rm CG}\pi}{3\ell^2}\left(a b \ell^2 + 6bn^2 + 12\hat{m}\right),\label{mass}
\end{align}
without need of boundary counterterms.

The particular mass value for Taub-NUT is obtained by replacing $\hat{m}=\hat{m}_{\rm nut}$ and $a=a_{\rm nut}$, while for Taub-Bolt by only replacing $\hat{m}=\hat{m}_{\rm bolt}$. Explicitly, we find
\begin{align}\label{Mnut}
     M_{\rm nut} &= 16\pi n\alpha_{\rm CG}\left[\frac{4(1-bn)}{\ell^2} - \frac{16n^2}{\ell^4}  -b^2\right]\,,\\
     \notag
     M_{\rm bolt} &= \frac{16\pi\alpha_{\rm CG}}{3}\bigg[ ab + \frac{6}{r_b\ell^4}\bigg\{r_b^4-6n^2r_b^2 - 3n^4 + \ell^2\Big[r_b^2+n^2 + a\left(r_b^2+n^2/3 \right) \\
     \label{Mbolt}
     &\quad +br_b\left(r_b^2+n^2 \right) \Big]  \bigg\} \bigg]\,.
\end{align}

The entropy, on the other hand, arises from different types of obstruction to foliation with hypersurfaces of constant time that the geometry might have~\cite{Hawking:1998jf,Garfinkle:2000ms,Ciambelli:2020qny}. For black holes, for instance, the entropy is obtained by evaluating the Noether charge at the horizon~\cite{Wald:1993nt}. In the case of Taub-NUT/Bolt, there is an additional contribution coming from the Misner string, giving
\begin{align}\label{entropy}
    S &= \beta\left[\int_0^{2\pi}\int_0^\pi\diff{\theta}\diff{\phi}\,Q_{\theta\phi}\Big|_{r=r_b} + \int_0^{2\pi}\int_{r_b}^\infty\diff{r}\diff{\phi}\,Q_{r\phi}\Big|_{\theta=\pi} - \int_0^{2\pi}\int_{r_b}^\infty\diff{r}\diff{\phi}\,Q_{r\phi}\Big|_{\theta=0} \right].
\end{align}
For Taub-NUT, the first term on the right hand side of Eq.~\eqref{entropy} vanishes identically since there is no horizon. Thus, the entropy comes purely from the Misner string, giving
\begin{align}\label{Snut}
    S_{\rm nut} = 32\pi^2\alpha_{\rm CG}\left[1-2b^2n^2 + \frac{8n^2(1-2bn)}{\ell^2}-\frac{48n^4}{\ell^4} \right].
\end{align}
The relevant thermodynamic variables for Taub-NUT satisfy the Gibbs-Duhem relation
\begin{align}
    S_{\rm nut} = 8\pi n\, M_{\rm nut} - I_{\rm nut},
\end{align}
and therefore the first law of thermodynamics is satisfied.

Additionally, when the Chern-Pontryagin invariant is added to the conformal gravity action  with fixed coupling $\vartheta=\pm16\pi^2\alpha_{\rm CG}$ [cf. Eq.~\eqref{ICGPont}], the Noether prepotential is modified according to
\begin{align}
    q_{\rm CGP}^{\mu\nu} &= -4\alpha_{\rm CG} \left[\left(W^{\mu\nu}_{\rho\sigma} \pm\tilde{W}^{\mu\nu}_{\rho\sigma} \right)\nabla^\rho\xi^\sigma + 2\xi^\rho\nabla^\sigma \left(W^{\mu\nu}_{\rho\sigma} \pm \tilde{W}^{\mu\nu}_{\rho\sigma}  \right) \right].
\end{align}
Thus, it is manifest that not only the Euclidean on-shell action vanishes when evaluated at the (anti-)self dual Taub-NUT solution, but their variations as well. This implies that introducing the Chern-Pontryagin topological term with fixed coupling constant to the conformal gravity action induce the triviality of conserved charges. 

 In Euclidean Einstein gravity, the thermodynamics of Taub-NUT/Bolt-AdS has been studied through different methods~\cite{Hawking:1998ct,Mann:1999pc,Emparan:1999pm,Astefanesei:2004ji,Mann:2004mi,Johnson:2014xza,Johnson:2014pwa,Ciambelli:2020qny}. In all cases, it has been observed that the Misner string not only contribute to the entropy but it diverges in AdS. This fact has motivated different renormalization schemes, ranging from background subtraction~\cite{Hawking:1998ct} to intrinsic boundary counterterms~\cite{Mann:1999pc,Emparan:1999pm}. Indeed, the addition of the Gauss-Bonnet term with fixed coupling renders the Misner string entropy finite~\cite{Ciambelli:2020qny} and it yields the action equivalent to the one proposed by MacDowell-Mansouri~\cite{MacDowell:1977jt}. For Einstein spaces, the latter reduces to Eq.~\eqref{ICG}, giving a consistent embedding of general relativity into conformal gravity modulo Neumann boundary conditions. Remarkably, the entropy of the Bach-flat Taub-NUT solution in Eq.~\eqref{Snut} coincides with the one obtained in Ref.~\cite{Ciambelli:2020qny} in the limit $b\to0$, provided a proper identification of the conformal gravity coupling, namely, $\alpha_{\rm CG}=\tfrac{\ell^2}{64\pi G}$. This, in turn, coincides with previous findings in the literature up to a thermodynamically irrelevant constant related to the Euler characteristic. 

\section{Comparison with generalized Taub-NUT metrics\label{sec:comparison}}

In this Section, we compare the Bach-flat solution presented in Eq.~\eqref{fsol2} with the metrics proposed in Refs.~\cite{IWAI199355,Miyake1995}. To do so, we first compare a particular curvature invariant between solutions to obtain the function that relates the radial coordinates of both spaces. Then, we compute a second curvature to test the equivalence of metrics. In particular, the line element used in Refs.~\cite{IWAI199355} is
\begin{align}\label{Iwai-Katayama}
    \diff{s^2} &= F(\rho)\left(\diff{\rho^2} + \rho^2 \diff{\Omega^2} \right) + G(\rho)\left(\diff{\psi} + \cos\theta\diff{\phi} \right)^2,
\end{align}
where $\diff{\Omega^2} = \diff{\theta^2} + \sin^2\theta\diff{\phi^2}$ and the metric functions are
\begin{align}\label{FyGIK}
    F(\rho) = \frac{\alpha}{\rho} + \beta \;\;\;\;\; \mbox{and} \;\;\;\;\; G(\rho) = \frac{\beta\rho^2+\alpha\rho}{\delta\rho^2+\gamma\rho+1},
\end{align}
where $\alpha,\beta,\gamma,\delta$ describe a four-parameter family of metrics. 

The relevant curvature invariants associated to this solution are
\begin{align}\notag
    R &= \frac{1}{{2(\beta\rho+\alpha)^3(\delta\rho^2+\gamma\rho+1)^2}}\bigg[ 6\alpha(\alpha \gamma-2\beta) + \left[3\alpha^2\delta^2 + 3\alpha \beta\gamma\delta - 3\beta^2\left(\gamma^2-\delta\right)\right]\rho^3 \\
    \label{RicciscalIK}
    &\quad + \left[9\alpha^2\gamma\delta - 3\beta\left(\gamma^2-4\delta\right)\alpha  - 9\beta^2\gamma\right]\rho^2 + \left[\left(3\gamma^2 + 15\delta\right)\alpha^2 - 9\beta\gamma\alpha  - 9\beta^2\right]\rho   \bigg]\,, \\
    \notag
    H_{\mu\nu}H^{\mu\nu} &= \frac{3\rho^2}{16\left(\beta\rho+\alpha  \right)^6\left(\delta\rho^2+\gamma\rho+1 \right)^4}\bigg[8\beta^2\delta^2(2\alpha \delta-\beta\gamma)^2\rho^6 + 8\beta\delta(2\alpha \delta-\beta\gamma)(2\alpha^2\delta^2 \\
    \notag
    &\quad + 4\alpha \beta\gamma\delta - \beta^2\gamma^2 - 6\beta^2\delta)\rho^5 + (9\alpha^4\delta^4 + 62\alpha^3\beta\gamma\delta^3 + 43\alpha^2\beta^2\gamma^2\delta^2 - 34\alpha \beta^3\gamma^3\delta  \\ 
    \notag 
    &\quad + 3\beta^4\gamma^4 - 118\alpha^2\beta^2\delta^3 - 122\alpha \beta^3\gamma\delta^2 + 50\beta^4\gamma^2\delta + 81\beta^4\delta^2)\rho^4 + (18\alpha^4\gamma\delta^3  \\  
    \notag
    &\quad + 64\alpha^3\beta\gamma^2\delta^2 + 16\alpha^2\beta^2\gamma^3\delta - 6\alpha \beta^3\gamma^4 - 80\alpha^3\beta\delta^3 - 116\alpha^2\beta^2\gamma\delta^2 - 120\alpha \beta^3\gamma^2\delta   \\  
    \notag 
    &\quad + 18\beta^4\gamma^3 + 112\alpha \beta^3\delta^2 + 106\beta^4\gamma\delta)\rho^3 + (19\alpha^4\gamma^2\delta^2 + 32\alpha^3\beta\gamma^3\delta + 7\alpha^2\beta^2\gamma^4     \\
    \notag 
    &\quad - 22\alpha^4\delta^3 - 76\alpha^3\beta\gamma\delta^2 - 104\alpha^2\beta^2\gamma^2\delta - 32\alpha \beta^3\gamma^3 + 148\alpha^2\beta^2\delta^2 + 20\alpha \beta^3\gamma\delta   \\
    \notag 
    &\quad + 43\beta^4\gamma^2 + 42\beta^4\delta)\rho^2 + (10\alpha^4\gamma^3\delta + 10\alpha^3\beta\gamma^4 - 22\alpha^4\gamma\delta^2 - 56\alpha^3\beta\gamma^2\delta  - 16\alpha^2\beta^2\gamma^3  \\
    \notag 
    &\quad + 80\alpha^3\beta\delta^2 + 28\alpha^2\beta^2\gamma\delta  -16\alpha \beta^3\gamma^2 + 16\alpha \beta^3\delta + 34\beta^4\gamma)\rho + 3\alpha^4\gamma^4 - 14\alpha^4\gamma^2\delta   \\
    \label{TLR2IK}
    &\quad - 2\alpha^3\beta\gamma^3 + 17\alpha^4\delta^2 + 6\alpha^3\beta\gamma\delta - 5\alpha^2\beta^2\gamma^2 + 10\alpha^2\beta^2\delta - 2\alpha \beta^3\gamma + 9\beta^4\bigg].
\end{align}

On the other hand, the metric ansatz used in Ref.~\cite{Miyake1995} is
\begin{align}\label{Miyake}
    \diff{s^2} &= F(\rho)\left[\left(\diff{\rho^2} + \rho^2 \diff{\Omega^2} \right) + G(\rho)\left(\diff{\psi} + \cos\theta\diff{\phi} \right)^2\right],
\end{align}
and it is sufficient for us to compare only with metrics of \cite{Miyake1995} that are non-Einstein, since the solution presented in Sec.~\ref{sec:BFTN} are of that sort. The first type of non-Einstein space of \cite{Miyake1995} is
\begin{align}\label{FyGM1}
    F(\rho) = 1+\beta\rho \;\;\;\;\; \mbox{and} \;\;\;\;\; G(\rho) = \left(\frac{\rho}{1+\beta\rho} \right)^2,
\end{align}
where $\beta$ is constant. The Ricci scalar and traceless Ricci squared associated to the latter are, respectively,
\begin{align}\label{RicciscalM1}
R &= -\frac{9}{2\rho^2(1+\beta\rho)}\,, \\
\label{TLR2M1}
H_{\mu\nu}H^{\mu\nu} &= \frac{67\beta^2\rho^2+78\beta\rho+27}{16\rho^4(1+\beta\rho)^4}.
\end{align}
The second non-Einstein space is given by
\begin{align}\label{FyGM2}
    F(\rho) = \left(\gamma-\rho^2 \right)\left(\gamma+2\beta\rho+\rho^2\right) \;\;\;\;\; \mbox{and} \;\;\;\;\; G(\rho) = \left[\frac{\rho\left(\gamma-\rho^2 \right)}{\gamma+2\beta\rho+\rho^2} \right]^2 \,,
\end{align}
where $\beta$ and $\gamma$ are constants defining a two-parameter family of metrics. Their curvature invariants are
\begin{align}\label{RicciscalM2}
    R &= -\frac{3\left(35\rho^4-34\gamma\rho^2+3\gamma^2\right)}{2\rho^2\left(\gamma-\rho^2\right)^3\left(\gamma+2\beta \rho+\rho^2\right)}\,, \\
    \notag
    H_{\mu\nu}H^{\mu\nu} &= \frac{1}{16\rho^4\left(\gamma-\rho^2 \right)^6\left(\gamma+2\beta\rho+\rho^2 \right)^4}\bigg[3675\rho^{12} + 10780\beta\rho^{11} + (9996\beta^2 - 2030\gamma)\rho^{10} \\
    \notag
    &\quad - 3444\beta\gamma\rho^9 - \gamma(10640\beta^2 - 4757\gamma)\rho^8 - 1384\beta\gamma^2\rho^7 + 12\gamma^2(422\beta^2 - 379\gamma)\rho^6 \\
    \notag 
    &\quad + 408\beta\gamma^3\rho^5 - \gamma^3(1424\beta^2 - 1365\gamma)\rho^4 + 12\beta\gamma^4\rho^3 + 2\gamma^4(134\beta^2 + 9\gamma)\rho^2 \\
    \label{TLR2M2}
    &\quad + 156\beta\gamma^5\rho+27\gamma^6 \bigg].
\end{align}

In order to differentiate among solutions, we first compare Eq.~\eqref{RicciscalBF} with Eqs.~\eqref{RicciscalIK},~\eqref{RicciscalM1}, and~\eqref{RicciscalM2}. This leads to a quadratic relation for $r$ in terms of $\rho$, that it can be solved analytically to find $r=r_\pm(\rho)$ in all the cases. We do not include the explicit solutions here because they cumbersome and not very illuminating. Afterward, we replace $r=r_\pm(\rho)$ into Eq.~\eqref{TLR2BF} and compare with Eqs.~\eqref{TLR2IK},~\eqref{TLR2M1}, and~\eqref{TLR2M2}. By doing so, we make sure that the radial coordinates among solutions are equivalent. Then, in the asymptotic region, the difference between solution becomes manifest since, as $\rho\to\infty$, we find
\begin{align}
    H_{\mu\nu}H^{\mu\nu}\Big|_{r=r_\pm(\rho)} \, = \, \varpi  + \mathcal{O}\left(\rho^{-3} \right) \,,
\end{align}
for the solution in Eq.~\eqref{fsol2}, where $\varpi$ is a non-vanishing constant that depends on the parameters of the solutions and whose explicit form is cumbersome and unimportant. In contrast, the traceless Ricci squared of Eqs.~\eqref{TLR2IK},~\eqref{TLR2M1}, and~\eqref{TLR2M2} behaves as $\mathcal{O}\left(\rho^{-\nu}\right)$ with $\nu\geq 6$ in all cases, as $\rho\to\infty$. This proves that the solution~\eqref{fsol2} is actually different from the ones in Refs.~\cite{IWAI199355,Miyake1995}. 

\section{Generalized Eguchi-Hanson instantons\label{sec:eguchi-hanson}}

So far, we have studied non-Einstein generalization of Taub-NUT/Bolt gravitational instanton solutions in conformal gravity. These are given by Bach-flat geometries that asymptote to locally maximally symmetric spacetimes, with fall off conditions that are in general weaker than those in Einstein gravity. This analysis can be easily extended to other solutions. In fact, other self-dual gravitational instantons in conformal gravity can be constructed: Consider, for example, the ansatz
\begin{align}
    \diff{s^2} &= \frac{r^2\,f(r)}{4}\left(\, \diff{\tau} +\, \cos\theta\, \diff{\phi}\, \right)^2 + \frac{\diff{r^2}}{f(r)} + \frac{r^2}{4}\left(\, \diff{\theta^2}\, +\, \sin^2\theta\, \diff{\phi^2} \right)\, ,\label{LaMetrica}
\end{align}
which is Bach-flat provided the function $f(r)$ is of the form
\begin{align}
    f(r) = 1 - \frac{a^4}{r^4}  + \frac{r^2}{\ell^2} + \frac{b}{r^2} + p\, r^4\,,\label{LaForma}
\end{align}
where $a,b,p,\ell$ are integration constants subject to the condition 
\begin{align}
    b = -4a^4p\ell^2.\label{ladearrioba}
\end{align}

The line element~\eqref{LaMetrica} with~\eqref{LaForma} is a generalization of the Eguchi-Hanson solution of Einstein gravity \cite{Eguchi:1978gw, Eguchi:1978xp}. The latter corresponds to the particular case $b=p=0$. In fact, the metric above can be written in the usual form \cite{Eguchi:1978gw, Eguchi:1978xp}; namely
\begin{eqnarray}
\diff{s^2} &=& \frac{dr^2}{f(r)} + r^2 \, (\, \sigma_1^2\, +\, \sigma^2_2\, +\, {f(r)}\, \sigma_3^2\,)
\end{eqnarray}
with the Maurer-Cartan forms of $SU(2)$ being defined as $\sigma_1= \frac 12  (\sin \tau \, d\theta - \sin\theta \, \cos \tau \, d\phi)$, $\sigma_2= \frac 12  (-\cos \tau \, d\theta - \sin\theta \, \sin \tau \, d\phi)$, and $\sigma_3= \frac 12  (d\tau +\cos \theta\, d\phi) $. Its scalar curvature is
\begin{align}
    R = -\frac{24}{\ell^2} - 48pr^2,
\end{align}
and it diverges as $r\to\infty$. To cure this, one could set $p=0$ that, provided $\ell^2$, fixes $b=0$. As in the case of asymptotically locally AdS sectors, the form of (\ref{LaForma}) permits to identify the Neumann boundary condition that selects the Eguchi-Hanson solution of general relativity out of the solutions of conformal gravity.


The (anti-)self duality condition, i.e. $W^{\mu\nu}_{\lambda\rho} = \pm \tilde{W}^{\mu\nu}_{\lambda\rho}$, of the generalized Eguchi-Hanson metric is achieved for particular values of the parameters. Specifically, we find that, once the condition~\eqref{ladearrioba} is imposed, the solution becomes self dual when $b=p=0$ and $\ell\to\infty$, reducing to the standard asymptotically locally flat Eguchi-Hanson metric of general relativity with $c=-3$ and $\chi=2$ as defined in Eqs.~\eqref{signatura} and~\eqref{eulerch}, respectively. 

There is another interesting possibility: first take $\ell\to\infty$ in Eq.~\eqref{LaForma} and then obtain the condition on the parameters imposed by the field equations $B_{\mu\nu}=0$. The latter yields $p=0$ and thus the metric becomes
\begin{align}
   f(r) = 1 - \frac{a^4}{r^4}  + \frac{b}{r^2} \,.\label{LaFormaAutodual}
\end{align}
This metric has been study in Refs.~\cite{Xiao_2004,Chen:2020org} and it is, indeed, (anti-)self dual, Bach-flat, and its Ricci scalar vanishes. However, it is non-Einstein as it can be seen from their traceless Ricci tensor squared
\begin{align}
    H_{\mu\nu}H^{\mu\nu} &= \frac{12 b^2}{r^8}\,.
\end{align}
This curvature invariant measures the deviation from solutions of general relativity since, recall, it vanishes identically for Einstein spaces. Therefore, it becomes clear that the $b$-mode renders the solutions non-Einstein. Reciprocally, this permits to identify the Newmann boundary condition that is necessary to impose on the conformal gravity solution to recover the Einstein solution in the Eguchi-Hanson sector with vanishing cosmological constant.

The absence of conical singularities at the bolt demands periodicity on the Euclidean time coordinate, namely $\tau\sim\tau+\beta_\tau$, where
\begin{align}
    \beta_\tau = \frac{2\pi}{1+\frac{b}{2r_b^2}}\,.
\end{align}
where the bolt radius is defined as $f(r_b) = 1 - a^4/r_b^4  + b/r_b^2 = 0$. The Chern-Pontryagin index and Euler characteristic of the solution~\eqref{LaFormaAutodual} are
\begin{align}\label{signaturaEH}
    c &=  -\left(3 + \frac{2b}{r_b^2} + \frac{b^2}{2r_b^4}\right)\left(1+\frac{b}{2r_b^2} \right)^{-1}  \,,\\  
    \chi &= 2  \,.
\end{align}
In the limit $b\to0$ these topological invariants reduce to those of the original Eguchi-Hanson metric, namely $c=-3$ and $\chi = 2$. These results for the topological numbers take into account the right periodicity of the angular variables, cf. \cite{Eguchi:1978gw}, and the normalization in~\eqref{signatura}. Other special cases are worth mentioning: When $a=0\neq b$ the Chern-Pontryagin index is $c=-3$ and $\chi=2$, which are equivalent to the invariants of the original Eguchi-Hanson metric.  

The Euclidean on-shell action of conformal gravity evaluated on the self-dual Bach-flat generalization of the Eguchi-Hanson solution~\eqref{LaMetrica} with metric function~\eqref{LaFormaAutodual} gives
\begin{align}
    I = \pm16\pi^2\alpha_{\rm CG}\, c\,,
\end{align}
with $c$ given in Eq.~\eqref{signaturaEH}. On the other hand, the Noether-Wald formalism can be used to compute the conserved charges. We find that the mass of the solution vanishes while the entropy, as defined in Eq.~\eqref{entropy}, gives
\begin{align}
    S = -8\pi\alpha_{\rm CG}\beta_\tau\left(3 + \frac{2b}{r_b^2} + \frac{b^2}{2r_b^4}\right) = -I\,.
\end{align}
This is consistent with the Gibbs-Duhem relation for a massless solution. Finally, we notice that adding the Chern-Pontryagin invariant with fixed coupling as given in Eq.~\eqref{Belucci} renders the Euclidean on-shell action and entropy of the generalized Eguchi-Hanson metric~\eqref{LaFormaAutodual} equal to zero. Thus, we conclude that this solution is part of the family of self-dual ground states of conformal gravity once topological terms are included. 

\section{Conclusions\label{sec:conclusion}}

Here, we have studied {non-Einstein} Bach-flat geometries that can be regarded as the generalization of single-centered gravitational instantons of Einstein gravity to conformal gravity. This includes generalizations of the Taub-NUT-Schwarzchild solution, the Euclidean version of the Riegert static black hole, and extensions of the Eguchi-Hanson metric. Both Taub-NUT and Taub-Bolt geometries were analyzed in detail. By computing the Chern-Pontryagin index and the Weil-Chern-Gauss-Bonnet integrals, we showed that the Bach-flat version of Taub-NUT/Bolt presents both non-vanishing Hirzebruch signature and Euler characteristic. The same holds for the generalizations of the Eguchi-Hanson metric.

For an arbitrary choice of the parameters, all the solutions we found generically present a curvature singularity at the origin, while at large distances they exhibit ALF or AlAdS asymptotics. Nevertheless, it is worth pointing out that the fall off behavior of the most general AlAdS solutions turns out to be weaker than the standard AdS$_4$ Henneaux-Teitelboim boundary conditions~\cite{Henneaux:1985tv}. This is associated to the presence of a low decaying mode of conformal gravity. In fact, our analysis permits to identify the simple Neumann boundary condition that, as observed by Maldacena in the asymptotically AdS sector \cite{Maldacena:2011mk}, selects the Einstein solution out of the solutions of conformal gravity.

When regularity at hypersurfaces of constant time is imposed, the existence of zero and two-dimensional fixed points---also referred to as nuts and bolts, respectively---yields to everywhere regular geometries; the former being a (anti-)self dual configuration that saturates the BPS bound of the Euclidean on-shell action. Indeed, we explicitly showed that, by adding the Chern-Pontryagin invariant to the conformal gravity action with fixed coupling constant, all (anti-)self dual configurations become the ground state of the theory, yielding vanishing free energy and conserved charges.

For the whole family of AlAdS Taub-NUT/Bolt and Eguchi-Hanson spaces, we computed the Noether charges, which, as usual in conformal gravity, happen to be finite without need of introducing counterterms. Also the entropy, the Euclidean action, and all the relevant quantities derived from it turn out to be finite. This enables us to study the thermodynamic properties of these geometries as well as it provides the necessary tools to perform semiclassical computations. We showed that the Gibbs-Duhem relation and the first law of thermodynamics are satisfied in all cases. This automatically guarantees the identity between the entropy obtained through the Noether-Wald formalism and the Euclidean methods.
 
Interesting questions remain open. For instance, the presence of additional gravitational hairs in conformal gravity portraits a richer scenario to study Hawking-Page phase transitions. Another question regarding the Euclidean approach is that of the finiteness of the conserved charges. Indeed, a detailed analysis of the relation between the conformal invariance and the finiteness of the Euclidean action in different sectors is worth pursuing. It would also be interesting to investigate the possible holographic interpretation of the massive $b$-mode of the AlAdS instantons studied here and their relation to the linear response in the conformal field theory at the boundary (see for instance Ref.~\cite{Ghodsi:2014hua}). The interpretation of low decaying modes, although in a different context, was discussed in~\cite{Gibbons:2006wd}. Last, it would be interesting to analyze the importance of the solutions we studied here in the context of topological gravity.

\begin{acknowledgments}
We thank G.~Anastasiou and I.~J.~Araya for insightful discussions and collaboration in early stages of this work. We also thank A.~Cisterna for valuable comments and for pointing out Ref.~\cite{Liu:2012xn} to us. G. G. thanks I. Lovrekovic for discussions. The work of C. C. is supported by Agencia Nacional de Investigación y Desarrollo (ANID) through FONDECYT No~11200025 and~1210500.  G. G. is supported by CONICET and ANPCyT grants PIP-1109-2017, PICT-2019-00303. The work of R. O. is funded by ANID through FONDECYT No~1170765. 
\end{acknowledgments}

\bibliography{References}

\end{document}